\documentclass[12pt]{article}
\usepackage{graphicx,epsfig,graphics,amssymb,amsmath,subeqnarray,color}

\def\O{\Omega}
\def\t{\bf t}\def\dl{\delta \ell}\def\d{\rm d}
\def\e{\epsilon} \def\c{\overline{c_{\parallel}}}\def\L{L_{\parallel}}
\def\MFO{{\cal M}^{F \O}}\def\MFU{{\cal M}^{F U}}\def\MLO{{\cal M}^{L \O}}\def\MLU{{\cal M}^{L U}}
\def\NFU{{\cal N}^{F U}}\def\NLO{{\cal N}^{L \O}}
\def\WLU{{\cal W}^{L U}}\def\WFU{{\cal W}^{F U}}\def\WFO{{\cal W}^{F \O}}\def\WLO{{\cal W}^{L \O}}
\def\VLU{{\cal V}^{L U}}\def\VFO{{\cal V}^{F \O}}
\def\V{{\cal V}}\def\W{{\cal W}}\def\M{{\cal M}}\def\N{{\cal N}}
\def\F{{\cal F}}\def\LL{{\cal L}}\def\I{{\cal I}}\def\J{{\cal J}}\def\R{{\cal R}}

\begin{document}

\title{\textbf{  \Large  Swimming in circles:  \\ Motion of bacteria near solid boundaries}}
\author{\normalsize  Eric Lauga$^{1,\dag}$, Willow R. DiLuzio$^{1,2,\ddag}$, George M. Whitesides$^{2,\dag\dag}$ \& Howard A. Stone$^{1,\ddag\ddag}$\\
\small \it $^1$Division of Engineering and Applied Sciences, \\
\small\it $^2$Department of Chemistry and Chemical Biology,\\
 \small \it Harvard University, Cambridge MA 02138.\\
\small  $^\dag$lauga@deas.harvard.edu, 
\small $^\ddag$wdiluzio@gmwgroup.harvard.edu, \\
\small $^{\dag\dag}$gwhitesides@gmwgroup.harvard.edu,
\small $^{\ddag\ddag}$has@deas.harvard.edu.
} 
\date{\normalsize \today}

\maketitle
\begin{abstract}
Near a solid boundary, {\it E. coli} swims in clockwise circular motion. We provide a hydrodynamic model for this behavior. We show that circular trajectories are natural consequences of  force-free and torque-free swimming, and the hydrodynamic interactions with the boundary, which also leads to a hydrodynamic trapping of the cells close to the surface. We compare the results of the model with  experimental data and  obtain reasonable agreement. In particular, we show that the radius of curvature of the trajectory increases with the length of the bacterium body.
\vskip2cm
\end{abstract}

\section{Introduction}

The bacterium {\it Escherichia coli} ({\it E. coli}) has been a micro-organism of choice for studying  biological and biomechanical processes. In particular, {\it E. coli} has been used as the prototypical  micro-swimmer \cite{berg00,bergbook}. 

{\it E. coli} and other peritrichously flagellated bacteria swim by the action of rotary motors (two to six) embedded in the cell wall.  All the motors rotate counter-clockwise or clockwise, when viewed from outside the cell, with each motor driving a long, thin, left-handed helical flagellar filament.  If all the motors rotate counter-clockwise in a viscous fluid (e.g. water), the flagella bundle together and propel the bacterial cell forward approximately in a straight line.  This motion is called a ``run". If one or more motors runs clockwise, the flagella unbundle, and the bacteria tumbles. The forward thrust generated by the flagellar bundle in a run is opposed by the translational viscous drag on the entire cell.  Each flagellum (average length, 5 to 7 $\mu$m) rotates  at speeds of approximately 100 Hz \cite{lowe87,magariyama01} and its counter-clockwise rotation exerts a net torque on the cell body (average length, $2$ to $5$ $\mu$m). To balance this torque, the cell body counter-rotates in a clockwise direction (viewed from behind the organism) at speeds of approximately 10 Hz \cite{macnab77}.  

When close to a solid surface,  {\it E. coli}  does not swim in a straight line, but  traces out a clockwise (when viewed from above the surface), circular trajectory \cite{maeda76,berg90,frymier95,frymier97,vigeant97,vigeant02}. The purpose of this paper is to provide a hydrodynamic model for such circular motion, examine how it depends on the  size of the cell and other cell parameters, and compare the results of the model to  experimental data.

An early observation of circular motion (1971), reported in \cite{berg90}, measured a radius of curvature for the circles on the order of 25~$\mu$m. The swimming direction was clockwise when viewed from above, which  the authors expected, as the flagellar bundle rotates counter-clockwise and the cell body rotates clockwise. The influence of temperature on the motility of {\it E. coli} was considered in \cite{maeda76}; this work reported circular curves for the motion near a glass slide, with a radius on the order of $10-50$~$\mu$m and increasing with  temperature. A tracking microscope was used later \cite{frymier95,frymier97} to follow the trajectory of {\it E. coli} near  a glass surface. Again, near solid boundaries, the bacteria were observed to swim in circles, with radius of about 13~$\mu$m; the authors also found that the  swimming speed  increased with the distance from the boundary. The question of attraction between the swimming bacteria and solid surface has  been studied in \cite{vigeant97,vigeant02}, and the distance that cells swim parallel to the surface has been measured (tens of nanometers). It was found that standard DLVO (Derjaguin-Landau-Verwey-Overbeek) theory could not explain the tendency of the cells to stay near the surfaces, but that some other force was still to be identified. The authors proposed that, because of their non-spherical shape, the cells swim at an angle to the surface, and therefore  constantly swim into the surface. 

Numerically, there has been only  one study that has considered the hydrodynamics of a swimming bacteria near a no-slip surface \cite{ramia93} (see also early work on flagellar motion near boundaries in \cite{reynolds65,katz74,katz75,katzblake,brennen77}). The bacterium was modeled as a body of  spherical shape with a single, solid, helical flagellum and the boundary integral method was used for the numerical investigation. In this approach,  the total flow field is given by  a distribution of fundamental singularities for Stokes flow along the surface of the  micro-organism. In the simulations, circular motion was obtained with a  radius of curvature on the order of the length of the micro-organism (10~$\mu$m), with a tendency for the micro-organism to swim towards the wall and ``crash'' into it. Furthermore, the authors proposed  a physical picture for a clockwise motion. However, no simple analytical model was proposed  and a numerical integration was required to obtain the cell trajectories.  

The goal of this paper is to provide a hydrodynamic model for the motion of {\it E. coli} near solid boundaries. In \S \ref{bacteria:exp}, we first summarize our experiments to obtain  a new set of data on swimming speed and circular trajectories for  {\it E. coli} strain HCB437 near solid surfaces. We  then present in \S\ref{bacteria:model} our geometrical model for {\it E. coli}, and the physical picture for the circular trajectory of the bacterium near a no-slip surface,  based on the change in hydrodynamic resistance of elements along the cell body due to the nearby surface. Using resistive-force theory, we calculate in \S\ref{bacteria:trajectory} the trajectory of the bacterium. Since the full model requires a matrix inversion to be evaluated, we also present an approximate analytical solution for the trajectory. In particular, we show that the circular motion is clockwise and  that the cells need to swim {into} the surface as a natural consequence of force-free and torque-free swimming. We illustrate the results of our two models (the full model and its analytical approximation), show their dependence on  various geometrical parameters of the cell, and compare the models with our  experiments  in \S\ref{bacteria:comparison}. We find that our models are consistent with experimental  swimming speeds and radii of curvature of the circular motions, and that they allow us to obtain an estimate for the relation between the size of the bacterium and its distance to the surface. The values of the various hydrodynamic mobilities used in the model are presented in Appendix \ref{bacteria:mob}, and the cell trajectory far from a surface is given in Appendix \ref{nowall}.

\begin{figure}[t]
\centering
\includegraphics[width=.85\textwidth]{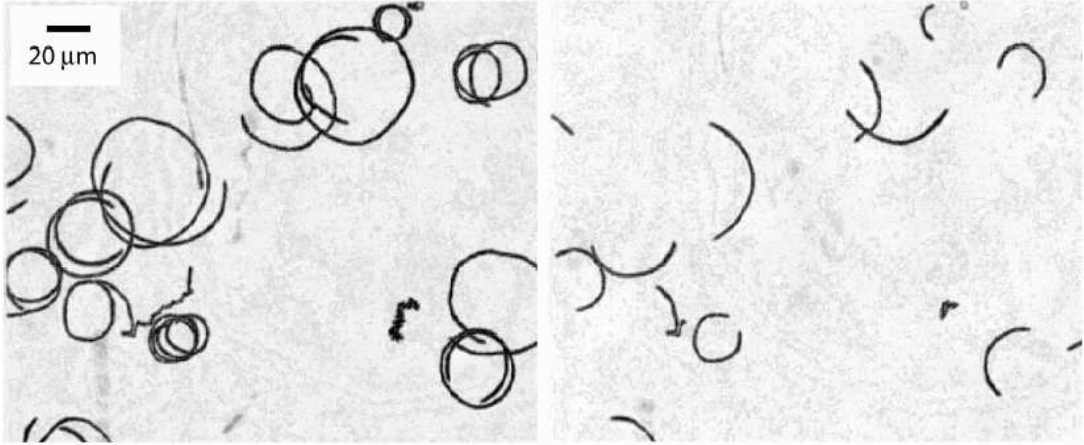}
\caption{Superimposed phase-contrast, video microscopy images show {\it E.coli} cells (HCB437) swimming in circular trajectories near a glass surface. Left:  Superposition of eight seconds (240 frames) of video images.  Right: Typical superposition of two seconds (60 frames) of video images that was used to analyze the length and width of cells, the swimming speed of cells, and the radius of curvature of the trajectories.} 
\label{typical}
\end{figure}

\section{Experiment}\label{bacteria:exp}
We examined a dilute suspension of smooth-swimming ({\it i.e.} non-tumbling) {\it E. coli} cells   (HCB437) \cite{wolfe87} in an observation chamber that was constructed from two glass cover-slips (separation, 80 $\mu$m) separated by double-sided tape.  Videos of cells swimming in circles below the top glass surface were collected, digitized, and analyzed.  The following parameters were tabulated for 90 individual cells and their trajectories over a two-second interval: Cell length and width, swimming speed, and the radius of curvature of the trajectory. 

The cells were observed from {\it outside} the chamber  above the surface, swimming with counter-clockwise trajectories; consequently, when viewed from  {\it within} the liquid (what we will refer to as ``above the surface'' in the remainder of the paper), they are performing clockwise trajectories.  In Fig.~\ref{typical} we provide superimposed video images  showing the curved trajectories that cells follow when swimming near the glass surface. 

\subsection{Materials and methods}
\label{method}
\subsubsection{Preparation of Motile Cells.}
{\it E. coli} strain HCB437 \cite{wolfe87} used in these studies  is a smooth-swimming strain that is deleted for most chemotaxis genes. During cell growth, cells double their length and then divide at their approximate midpoint (septate), while maintaining a constant width. The length of cells naturally vary depending on the progress of cells through the growth cycle \cite{growth}.  Media components were purchased from Difco or Sigma.  Saturated {\it E. coli} cultures were grown for 16 hours in tryptone broth (1\% tryptone and 0.5\% NaCl) using a rotary shaker (200~rpm) at 33 $^\circ$C.  Saturated cultures were frozen at $-70$ $^\circ$C in 15\% glycerol.  Motile {\it E. coli} cultures were obtained by diluting 50~$\mu$L of the thawed saturated culture into 5~mL of fresh tryptone broth, and grown in 14~mL sterile, polypropylene tubes at 33 $^\circ$C on a rotary shaker (150~rpm) for 3.5 hours.  Cells were washed by three successive centrifugations at 2000G for 8 minutes and were resuspended into motility buffer  \cite{adler67} (1~ mM potassium phosphate, pH~7.0, 0.1~mM Na-EDTA) containing 10~mM glucose and 0.18\% (w/v) methylcellulose (Methocel 90). Glucose was added to maintain motility in an anaerobic environment and methylcellulose was added to reduce the tendency of cells to wobble \cite{berg1972}.  Filamentous cells were obtained by growing motile cells for 3.5 hours as described above, adding 50~$\mu$g/mL cephalexin to the culture, and then growing cells an additional 0.5 hours  \cite{maki00}.  Filamentous cells were then washed as described above.

\begin{figure}[t]
\centering
\includegraphics[width=.99\textwidth]{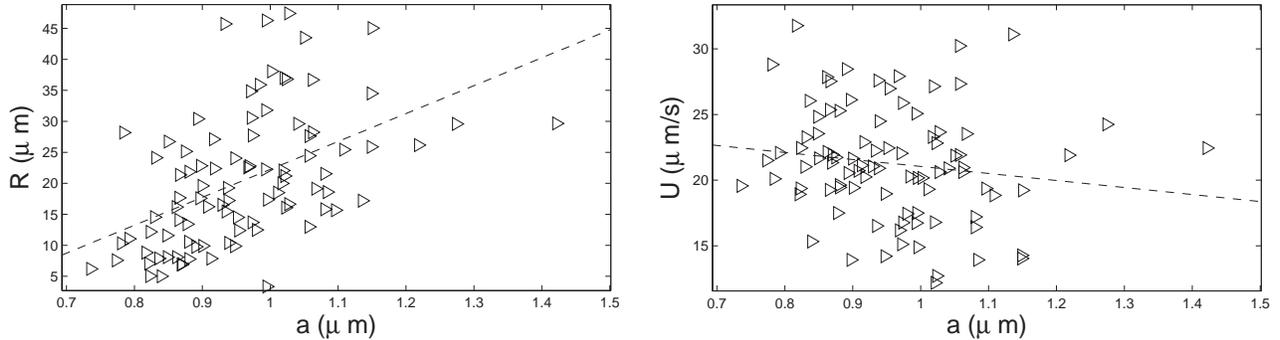}
\caption{Results of our experimental investigation of swimming {\it E. coli} near solid boundaries. Left:
Radius of curvature of the circular trajectory, ${\cal R}$,
 as a function of the equivalent sphere radius, $a$, of the elliptical cell body (see text); 
Right:  Swimming speed, $U$,  versus equivalent body radius, $a$. In both cases, we have added the best least-square fit to the data (dashed line).}\label{data}
\end{figure}

\subsubsection{Observation of Swimming Cells.}
A volume of 50~$\mu$L of the washed cell suspension (approximately $10^7$~cells/mL) was added to an observation chamber constructed from two glass coverslips and double-sided tape (Scotch, Permanent). The chamber dimensions were approximately: 1~cm wide, 2~cm long, and 80~$\mu$m high.  The microscope coverslips were alternately rinsed with soap and DI water, DI water, ethanol, DI water, and then treated with an air plasma for 1 minute at 1-2~Torr (SPI Plasma Prep II, power $\approx$80\%).  The observation chamber was heated to 32 $^\circ$C using a heated microscope stage (Research Instruments Limited). Cells swimming near the upper glass coverslip were observed using a Nikon Eclipse E400 upright, phase-contrast microscope.   Video images were acquired using a 20$\times$ or 40$\times$ Nikon phase objective and a monochrome CCD camera (Marshall Electronics V1070) connected to a digital video recorder (Sony GV-D1000) that collected 640 pixel $\times$ 480 pixel images at 30 frames per second.  

\subsubsection{Image Analysis.}
Video was captured into a computer using Adobe Premiere and analyzed using ImageJ (available for download at {http://rsbweb.nih.gov/ij/}) or Scion Image (available for download at {http://www.scioncorp.com}) using standard analysis tools.  Video images were thresholded so that cells appeared black and the background appeared white.  The following parameters were measured for individual cells in 60 consecutive video frames (2 seconds):  The projected area of the cell, the midpoint of the cell, and the short and long axis of the cell (approximating the cell shape as an ellipse).  The average of these values measured over the 2 s interval was used.  The average cell speed was calculated by measuring the average distance that the midpoint of the cell traveled between each video frame and dividing this distance by the video collection rate (30~fps or 0.033~s).  The radius of curvature of the cell trajectory was calculated by making  a least square fit of a circle to the two-second trajectory of the midpoint of the cell.  A small amount of error was introduced by the collection and analysis of cells from multiple regions of the swimming chambers and from multiple chambers.  Small changes in focus in different regions led to variability in the thresholding, which led to some error in the measurement of cell widths and lengths.  The variability in the cell widths was $0.98\pm.09$~$\mu$m; therefore, the error due to these small focus changes was less than 10\%.

\subsection{Results}

 In Fig.~\ref{data} we plot the experimental results for the cell swimming speed ($U$) and the radius of curvature of the circles (${\cal R}$) as a function of the equivalent sphere radius, $a$, that is, the radius of the sphere that has the same viscous resistance as the prolate ellipsoid of measured cell dimensions translating along its axis of symmetry \cite{happel}.  The large scatter in the experimental data, evident in Figs.~\ref{typical} and \ref{data}, can be explained by the natural cell-to-cell variability in the number of flagella \cite{turner00}, flagellar rotation rates \cite{magariyama01}, and distances of cells from the surface \cite{vigeant02} (parameters that might also be function of cell length).  Nonetheless, the experimental data demonstrate that the radius of curvature of the trajectories of cells  swimming near a glass surface increases with the length of the cell body.

\section{Model}\label{bacteria:model}
We present in this section our hydrodynamic model for the motion of {\it E. coli} near a flat no-slip surface and give a simple physical picture for the circular trajectory.

\subsection{Set-up}
\begin{figure}[t]
\centering
\includegraphics[width=.7\textwidth]{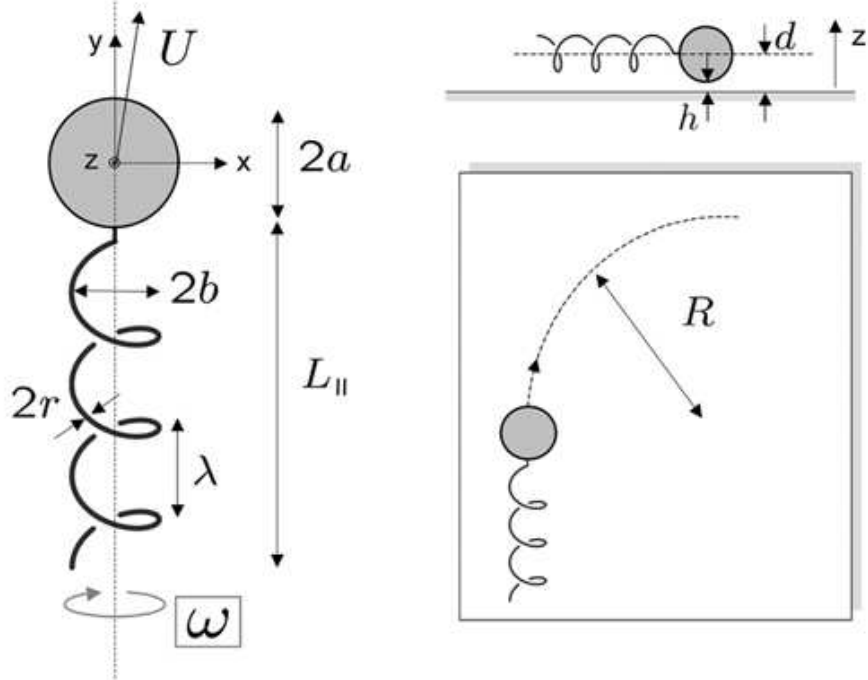}
\caption{Set-up and notations for the mechanical model of {\it E. coli} swimming near a solid surface.} \label{setup}
\end{figure}

We model the bacterium as a single, left-handed rigid helix attached to a spherical body of radius $a$ whose center of mass is located at a distance $d$ above a solid surface, as illustrated in Fig.~\ref{setup}; the liquid gap between the solid surface and the cell body has height $h$. The cell is assumed to be parallel to the surface and oriented in the $y$-direction. The helix is assumed to have  thickness $2r$, radius $b$, wavelength $\lambda$, with a number $n$ of wavelengths along the helix, such that the total length of the helix along the $y$-direction is $\L=n\lambda$. The assumption of sphericity, although not completely realistic for the cell body of {\it E. coli}, which is more like a 2:1 prolate ellipsoid, was made in order to use well-known mobility formulae, and we expect therefore our results  to be correct within a shape factor of order unity. Due to the action of rotary motors, the bundle is rotating in the counter-clockwise direction (viewed from behind) with an  angular velocity $\boldsymbol{\omega}=- \omega\, {\bf e}_y$  relative to the body, with $\omega>0$ (see Fig.~\ref{setup}). We denote by ${\bf U}=(U_x,U_y,U_z)$ and $\boldsymbol{\Omega}=(\O_x,\O_y,\O_z)$ the instantaneous velocity and rotation rate (measured from the center of the cell body), respectively, of the  bacterium.

\subsection{Physical picture}\label{picture}

\begin{figure}[t]
\centering
\includegraphics[width=.9\textwidth]{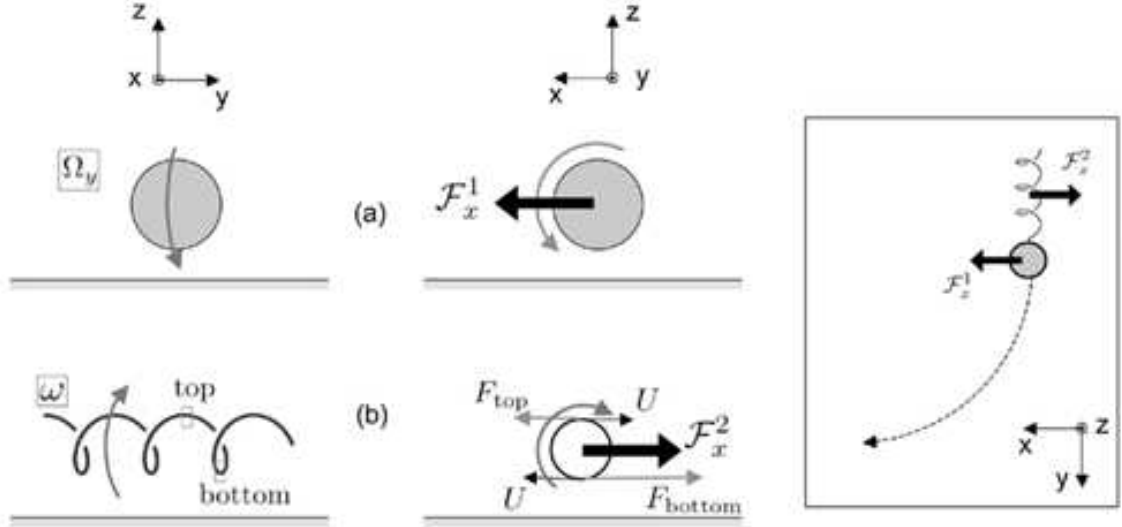}
\caption{Physical picture for the out-of-plane rotation of the bacterium: (a) The positive $y$-rotation of the cell body  leads to a net viscous $x$-force on the cell body, $\F_x^1>0$; (b) The negative $y$-rotation of the helical bundle leads to a net negative viscous $x$-force on the flagella, $\F_x^2 <0$. The spatial distribution of these forces leads to a negative $z$-torque on the bacterium, which makes it rotate clockwise around the $z$-axis. Therefore, when viewed from above, the bacterium swims to its right.} \label{rotate}
\end{figure}

In the absence of a nearby wall, the bacterium swims approximately in a straight line, ${\bf U}=U_y {\bf e}_y$, and rotates along its swimming axis, $\boldsymbol{\Omega}=\O_y{\bf e}_y$. The velocity $U_y>0$ is obtained by balancing the propulsive force of the helical bundle with the viscous resistance on the whole bacterium and the rotation rate $\O_y>0$ is found by the balance of viscous moments about the $y$-axis (see Appendix \ref{nowall}). 

What changes when the micro-organism is swimming near a solid surface? The cell body and the helical bundle contribute together to a rotation of the bacterium around the $z$-axis (see notation in Fig.~\ref{setup}) (see also \cite{yates86}).  First, as the cell body is near the surface, when it rotates about the $y$-axis at a rate $\O_y>0$, there is a viscous force acting on the cell body in the $x$-direction, $\F_x^1{\bf e}_x$, with  $\F_x^1>0$ (this is a standard hydrodynamic result, see \cite{cox70}).  Furthermore, the bundle of flagella is also acted upon by a net force in the $x$-direction. Since the bundle takes the shape of a helix, parts of the bundle are located close to the surface and parts are located further away. The local drag coefficient on an elongated filament decreases with increasing distance to the nearby surface  (see details below), which means that the parts of the bundle that are close to the surface will be subjected to a larger viscous force compared to  portions of the helix located further away from the surface. As the bundle rotates counter-clockwise around the $y$-axis (viewed from behind), the portions of the helix that are closer to the surface have a positive $x$-velocity, and therefore the net viscous force acting on the bundle, $\F_x^2{\bf e}_x$, is negative,  $\F_x^2<0$. Of course, since the swimming bacterium as a whole is force-free, we have necessarily $\F_x^2 = - \F_x^1$. 

As a consequence of the viscous forces acting on both the helical bundle and the cell body and their spatial distribution, a negative torque, $\LL_z <0$, will act on the bacterium and will rotate the entire cell clockwise around the $z$-axis (Fig.~\ref{rotate}). When viewed from above, the bacterium will therefore  swim to the right, as is observed experimentally. Note that, since the bacterium as a whole is torque-free (the Reynolds number is low, $Re\approx 10^{-4}$, so forces and torques need to balance at each instant), this torque will be balanced by a positive torque arising from the viscous resistance to a rotation around the $z$-axis. 

This physical picture  allows us to obtain an estimate for the radius of curvature $\R$ of the motion, as the ratio of the swimming velocity $U_y$ to the out-of-plane rotation rate $\O_z$. Since the Reynolds number is low, the equations of motion for the fluid are linear (Stokes flow) and therefore instantaneous viscous forces and torques for various parts of the bacterium are linearly related to their velocities and rotation rates. Let us denote by $\M$ and $\N$ the viscous (tensorial) mobilities of the bacterium flagella and body respectively which are non-zero away from a solid surface, and denote by $\W$ and $\V$ those mobilities that arise due to the presence of the nearby surface and tend to zero far from a wall (see \S \ref{bacteria:trajectory} for details). For all these mobilities, we will use notations of the form $\M^{\alpha\beta}_{ij}$, where the superscript $\alpha\beta$ is either $FU$ (in which case $\M^{FU}_{ij}$ denotes how the $i$th component  of a viscous force is linearly related to the $j$th component  of a velocity), $F\Omega$ (force - rotation rate), $LU$ (torque - velocity) or $L\Omega$ (torque - rotation rate). We also always use the convention that the mobilities are positive, and might therefore appear with a minus sign when necessary.

The swimming velocity is obtained by balancing the propulsive force of the micro-organism, $\MFO_{yy}(\omega-\O_y)$, with the viscous drag on the bacteria $(\MFU_{yy}+\NFU_{yy})U_y$, so that
\begin{equation}
\MFO_{yy}(\omega-\O_y) \approx (\MFU_{yy}+\NFU_{yy})U_y.
\end{equation}
The rotation rate can be estimated by balancing the wall-induced torque  mentioned above, $\LL_z \approx \WLO_{zy}(\omega-\O_y)$, with the viscous resistance to rotation of the whole bacterium, which is mostly due  to the viscous resistance of the flagella, $-\MLO_{zz}\O_z$, that is
\begin{equation}
\WLO_{zy}(\omega-\O_y) \approx -\MLO_{zz}\O_z.
\end{equation}
The two previous balances lead to an estimate for the radius of the circular motion as 
\begin{equation}\label{R_model}
{\cal R}\approx \frac{U_y}{|\Omega_z|} \approx \frac{\MLO_{zz}\MFO_{yy}}{\WLO_{zy}(\MFU_{yy}+\NFU_{yy})}\cdot 
\end{equation}
As is demonstrated below, this simple estimate is consistent with a more detailed calculation, as well as with experimental results.

\section{Trajectory calculation for the bacterium}\label{bacteria:trajectory}
We proceed in this section by presenting the detailed calculation for the trajectory of the bacterium using resistive-force theory for the flagellar hydrodynamics, and  exploit it  to obtain an approximate analytical solution.

\subsection{Modeling of flagella hydrodynamics }\label{modeling} 

The modeling chosen here for the helical hydrodynamics is that of resistive-force theory (RFT), as first introduced by Gray and Hancock \cite{gray55}, since it is the simplest approach to the zero-Reynolds-number hydrodynamics of elongated bodies. The method is an approximation to the equations of slender-body theory (SBT). SBT considers the zero-Reynolds-number dynamics of long and slender filaments by distributing fundamental Stokes flow singularities at their centerline (Stokeslet and force-dipole) \cite{cox70,keller76}. The idea was first introduced by Hancock \cite{hancock53}, is reviewed in detail by Lighthill \cite{lighthill76}, and has been applied to the case of helical flagella in \cite{higdon79}.  

RFT is the leading-order approximation of SBT, which gives results accurate at order ${\cal O}([\log(L/r)]^{-1})$, where $L$ is the length along the filament and $r$ its radius; the next term in the formulation of SBT appears at order ${\cal O}(1)$. The complexity of fully solving for the spatial distribution of singularities on a moving flagellar filament is replaced by introducing a set of local drag coefficients. Let us consider a  portion of the filament of length $\dl$, oriented along the tangential vector, $\t$, and moving at a velocity ${\bf u}$ in a viscous liquid. The local velocity can be decomposed into a parallel and perpendicular components, ${\bf u}={\bf u}_{\parallel}+{\bf u}_{\perp}$, where ${\bf u}_{\parallel}$ is parallel to the tangential vector, ${\bf u}_{\parallel}= ({\bf u}\cdot \t)\t$, and ${\bf u}_{\perp}$ is perpendicular to it, ${\bf u}_{\perp}={\bf u}-{\bf u}_{\parallel}$. RFT assigns values for the local drag coefficients, $c_{\parallel}$ and $c_{\perp}$, which relate the local viscous force per unit length to the local parallel and perpendicular velocities, such that the total force on an element of length $\delta \ell $ can be written
\begin{equation}
\delta {\bf F} =- \delta \ell( c_{\parallel}{\bf u}_{\parallel} +c_{\perp} {\bf u}_{\perp}).
\end{equation}
For a periodic flagellar filament (wavelength $\lambda$) performing planar oscillations  far from a solid surface, we have  approximately \cite{gray55,brennen77}
\begin{equation}
c_{\parallel}=\frac{2\pi\mu}{\ln(2\lambda/r)-1/2},\quad c_{\perp} = 2 \,c_{\parallel}\cdot
\end{equation}
The case of helical flagella was first considered in this context in \cite{chwang71}. Note that the drag anisotropy between tangential and perpendicular motion is the fundamental origin of the flagellar propulsion of micro-organisms \cite{bergbook,gray55,lighthill76}. Although it is only an approximate method, RFT has been shown in the past to provide both qualitative and quantitative information about the locomotion of micro-organisms \cite{brennen77,gray55,lighthill76,childress81,wiggins98}.

The presence of a solid surface modifies the values of the resistance coefficients for both the cell body and its flagella \cite{jeffrey,goldman67,oneill67,cooley68,jeffrey81,brennen77,katz75,katzblake,katz74}. Elements of the helical flagella are located at a distance $d(z)$ ranging between $d-b$ and $d+b$  to the solid surface, which are both  smaller than the helix wavelength $\lambda$, so that the viscous resistance  to motion of the flagella is dominated by the interactions with the surface.
Since $r\ll d\pm b$, we consider the far-field asymptotic results of \cite{katz75} (see also the review in \cite{brennen77}) and 
use 
\begin{equation}\label{drag}
c_{\parallel}(z) = \frac{2\pi \eta}{\ln\left({2d(z)}/{r}\right)},\quad c_{\perp} = 2\, c_{\parallel}\cdot
\end{equation}
Deviations from 2 for the  ratio $c_{\perp}/c_{\parallel}$  were discussed in this context by Katz \& Blake \cite{katzblake}. We will denote by  $\c$ the value of the drag coefficient, Eq.~\eqref{drag}, when  $d(z)=d$, and will denote deviations from this value by the function $f$, so that ${c_{\parallel}}(z)={\c}f(z)$.

\subsection{Mobilities}

We consider separately the mobilities of the cell body and its flagella, neglecting therefore the hydrodynamic interactions between these two parts of the micro-organism. Although this an approximation, we expect it will contribute only to a small error in the final results as the presence of a nearby surface leads to spatially localized flow fields, decaying at least as fast as a Stokeslet-dipole ($\sim 1/r^2$).

As described earlier, we denote by $\M$ and $\N$ the mobilities that are non-zero even in the absence of a wall, and by $\V$ and $\W$ those which are equal to zero when the micro-organism swims far from the surface (with the conventions that the mobilities are positive). The mobility matrix for the spherical cell can be written as
\begin{equation}\label{head_mobilities}
\left(
\begin{array}{c}
{\cal F}_x  \\
{\cal F}_y  \\
{\cal F}_z  \\
{\cal L}_x  \\
{\cal L}_y  \\
{\cal L}_z
\end{array}
\right)=
\underbrace{
\left(
\begin{array}{cccccc}
 -\NFU_{xx}  &  0   &   0   & 0  & \VFO_{xy}  & 0 \\
 0   &  -\NFU_{yy}   &   0   & -\VFO_{yx}   & 0  & 0 \\
 0   &  0   &   -\NFU_{zz}    & 0  & 0  & 0 \\
 0   &  -\VLU_{xy}   &  0   & -\NLO_{xx}  & 0  & 0 \\
 \VLU_{yx}    &  0   &  0   & 0  & -\NLO_{yy}  & 0 \\
 0   &  0   &  0   & 0  & 0  & -\NLO_{zz} \\
\end{array}
\right)
}_{\displaystyle \cal A}
\cdot
\left(
\begin{array}{c}
U_x  \\
U_y  \\
U_z  \\
\O_x  \\
\O_y  \\
\O_z
\end{array}
\right),
\end{equation}
and that of the helical flagella as
\begin{equation} \label{flagella_mobilities}
\left(
\begin{array}{c}
{\cal F}_x  \\
{\cal F}_y  \\
{\cal F}_z  \\
{\cal L}_x  \\
{\cal L}_y  \\
{\cal L}_z
\end{array}
\right)=
\underbrace{\left(
\begin{array}{cccccc}
 -\MFU_{xx}  &   \WFU_{xy}   &   0   & \MFO_{xx}  & \WFO_{xy}  & -\MFO_{xz} \\
 \WFU_{yx}   &  -\MFU_{yy}  &   0   & -\WFO_{yx}  & -\MFO_{yy}   & \WFO_{yz} \\
 0   &  0   &  -\MFU_{zz}   & \MFO_{zx}   & 0  & \MFO_{zz} \\
 \MLU_{xx}   &  - \WLU_{xy}   &  \MLU_{xz}  & -\MLO_{xx}  & -\WLO_{xy}  & 0 \\
 \WLU_{yx}  &  \MLU_{yy}    &  0   & -\WLO_{yx} & -\MLO_{yy}  & \WLO_{yz} \\
  -\MLU_{zx}  &  \WLU_{zy}   &  \MLU_{zz}   & 0  & \WLO_{zy}   & -\MLO_{zz} \\
\end{array}
\right)}_{\displaystyle \cal B}
\cdot
\left(
\begin{array}{c}
U_x  \\
U_y  \\
U_z  \\
\O_x  \\
\O_y-\omega  \\
\O_z
\end{array}
\right),
\end{equation}
with values  calculated in Appendix \ref{bacteria:mob}. As can be seen in Eq.~\eqref{flagella_mobilities}, 
the matrix ${\cal B}$  is almost full; the elements reported to be zero are either exactly zero at all instants or   time-average to zero over the rotation period of a flagellar filament, $T=2\pi/\omega$.
If we define 
\begin{equation}
{\cal X}=(U_x, U_y ,U_z, \O_x, \O_y, \O_z)^T, \quad {\rm and}\quad {\cal Y}=(0,0,0,0,\omega,0)^T,
\end{equation}
then the requirement that the micro-organism is free-swimming,
${\boldsymbol{\cal L}}={\bf 0}$ and $ {\boldsymbol{\cal F}}={\bf 0}$,
becomes a 6$\times$6 linear system to solve for ${\cal X}$ of the form
\begin{equation}\label{tosolve}
({\cal A}+{\cal B}) {\cal X}={\cal B}{\cal Y}.
\end{equation}
The solution for the velocity, ${\bf U}$, and the rotation rate, $\boldsymbol{\Omega}$, can be found by simply substituting the 
values of the mobilities from Appendix \ref{bacteria:mob} and solving the linear system Eq.~\eqref{tosolve}. 
The radius of curvature of the in-plane motion will then be given by 
\begin{equation}
{\cal R}= \frac{U}{|\Omega_z|},\quad U=\sqrt{U_x^2+U_y^2}\cdot
\end{equation}

\subsection{Approximate analytical solution}\label{bacteria:approx}

When the bacteria swims far from the solid surface, an analytical solution to motion can be found and we give it in Appendix \ref{nowall}. In the presence of a solid surface, an analytical solution to the linear system, Eq.~\eqref{tosolve}, exists in theory by direct matrix inversion, but it is very complicated and not very enlightening. We present below instead an approximate analytical solution of the linear system.

First, we note that,  in the case of {\it E. coli}, a number mobilities can be neglected between the elements of ${\cal A}$ and ${\cal B}$. They are
\begin{subeqnarray}
\NLO_{zz}\ll\MLO_{zz},&& \NLO_{xx}\ll\MLO_{xx},\\
\MFU_{zz}\ll \NFU_{zz},&&\MLO_{yy}\ll \NLO_{yy},\\
\WLU_{xy}\ll \VLU_{xy}, & & \WLU_{yx}\ll\VLU_{yx},\\
\WFO_{xy}\ll \VFO_{xy}, && \WFO_{yx}\ll \VFO_{yx}.
\end{subeqnarray}

Furthermore, since the $x$- and $z$-components of both velocity and rotation rate are zero far from the solid surface, we make the assumption that, near the surface, these components are at most on the order of the $y$-components: we therefore assume that  $(U_x,U_z) \lesssim U_y$ and $(\O_x, \O_z) \lesssim \O_y$. We further assume that $\O_y \ll \omega$, as is the case far  from the surface. Finally, since we have in general $ ({\cal W}_{ij}^{\alpha\beta} ,{\cal V}_{ij}^{\alpha\beta} )\ll (\M^{\alpha\beta}_{iy},\N^{\alpha\beta}_{iy})$, where $j=x$ or $z$, 
these assumptions allow us to simplify further the mobilities in the matrices ${\cal A}$ and ${\cal B}$.

In that case, the equations $\sum\LL_y=0$ and $\sum\F_y=0$ lead to the approximate solutions for the swimming speed and body rotation
\begin{subeqnarray}
\slabel{Uy_approx} U_y & \approx & \frac{\MFO_{yy}}{\MFU_{yy}+\NFU_{yy}}\,\omega, \\ 
\O_y & \approx & \frac{\MLU_{yy}\MFO_{yy}}{\NLO_{yy}(\MFU_{yy}+\NFU_{yy})}\,\omega. \slabel{Oy_approx}
\end{subeqnarray}
and $\O_y$ is indeed verified to be much smaller than $\omega$. We can then use $\sum\F_z=0$ and obtain
\begin{equation}\label{Uz_mid}
U_z=\frac{1}{\NFU_{zz}}\left[
\MFO_{zx}\O_x+\MFO_{zz}\O_z
\right].
\end{equation}
It follows, by substituting Eq.~\eqref{Uz_mid} into $\sum\LL_z=0$ and evaluating the leading-order contribution, that
\begin{equation}\label{Ux_mid}
U_x=\frac{1}{\MLU_{zx}}\left[
\frac{\MLU_{zz}\MFO_{zx}}{\NFU_{zz}}\O_x-\MLO_{zz}\O_z-\WLO_{zy}\omega
\right].
\end{equation}
As a consequence, substituting Eqs.~\eqref{Uz_mid} and  \eqref{Ux_mid} into $\sum\LL_x=0$, using Eq.~\eqref{Uy_approx} and evaluating the leading-order term leads to
\begin{equation}\label{eq1}
\MLO_{xx}\O_x+\frac{\MLO_{zz}\MLU_{xx}}{\MLU_{zx}}\O_z+\frac{\VLU_{xy}\MFO_{yy}}{\MFU_{yy}+\NFU_{yy}}\omega=0.
\end{equation}
Finally, substituting Eqs.~\eqref{Uy_approx}, \eqref{Oy_approx} and \eqref{Ux_mid} into $\sum\F_x=0$ and keeping the leading-order terms leads to 
\begin{equation}\label{eq2}
\MFO_{xx}\O_x+\left[\frac{\MLO_{zz}(\MFU_{xx}+\NFU_{xx})}{\MLU_{zx}}-\MFO_{xz} \right]\O_z+\frac{\WLO_{zy}(\MFU_{xx}+\NFU_{xx})}{\MLU_{zx}}\omega=0.
\end{equation}
Solving the $2\times 2$ linear systems of equations given by  Eqs.~\eqref{eq1}-\eqref{eq2}, 
and keeping only the leading-order terms leads to approximate formulae for the $x$- and $z$-components of the rotation rates as
\begin{subeqnarray}
\slabel{Ox_approx} \O_x & \approx & -\frac{\VLU_{xy}\MFO_{yy}}{\MLO_{xx}(\MFU_{yy}+\NFU_{yy})}\,\omega, \\
\slabel{Oz_approx} \O_z & \approx & -\frac{\WLO_{zy}}{\displaystyle \left(\MLO_{zz}-\frac{\MFO_{xz}\MLU_{zx}}{\MFU_{xx}+\NFU_{xx}}\right)}\,\omega.
\end{subeqnarray}
Note that the denominator in the equation for  $\O_z$, Eq.~\eqref{Oz_approx}, is dominated by $\MLO_{zz}$ but not by much, so we need to keep both terms to obtain correct orders of magnitude. These equations allow us to verify that, for {\it E. coli}, $\O_x$ is  much smaller than $\O_y$ and $\O_z$ is of the same order as $\O_y$. Note also that we obtain $\O_z < 0$, which means that the bacteria is swimming to its right (clockwise trajectory viewed from above) and that $\O_x < 0$, so that the bacteria will also have the tendency to swim into the surface. Also, we observe that
\begin{equation}
\frac{a\O_x}{U_y}\approx \frac{a\VLU_{xy}}{\MLO_{xx}}\approx \left(\frac{a}{\L}\right)^3\ll 1,
\end{equation}
so the time scale for re-orientation of the bacteria perpendicular to the surface is 
much larger than the typical swimming time scale.

 \begin{table}[t]
\begin{center}
\begin{tabular}{ll}
\hline 
\\
$\displaystyle U_x \approx  \frac{\displaystyle\WLO_{zy}}{\displaystyle \left(\frac{\MLO_{zz}(\MFU_{xx}+\NFU_{xx})}{\MFO_{xz}}-\MLU_{zx}\right)}\,\omega $
&
 $\displaystyle\O_x  \approx  -\frac{\VLU_{xy}\MFO_{yy}}{\MLO_{xx}(\MFU_{yy}+\NFU_{yy})}\,\omega$\\ \\
$\displaystyle U_y  \approx  \frac{\MFO_{yy}}{\MFU_{yy}+\NFU_{yy}}\,\omega$ 
& 
$\displaystyle \O_y  \approx  \frac{\MLU_{yy}\MFO_{yy}}{\NLO_{yy}(\MFU_{yy}+\NFU_{yy})}\,\omega$\\ \\
$\displaystyle U_z  \approx -\frac{\VLU_{xy}\MFO_{zx}\MFO_{yy}}{\NFU_{zz}\MLO_{xx}(\MFU_{yy}+\NFU_{yy})}\,\omega$ 
& 
$\displaystyle \O_z  \approx  -\frac{\WLO_{zy}}{\displaystyle \left(\MLO_{zz}-\frac{\MFO_{xz}\MLU_{zx}}{\MFU_{xx}+\NFU_{xx}}\right)}\,\omega$\\ \\
 $\displaystyle{\cal R}  \approx  \frac{\MLO_{zz}\MFO_{yy}}{\WLO_{zy}(\MFU_{yy}+\NFU_{yy})} \left(1-\frac{\MFO_{xz}\MLU_{zx}}{\MLO_{zz}(\MFU_{xx}+\NFU_{xx})}\right)$& \\
\\
\hline
\end{tabular}
\end{center}
\caption{Summary of the results of the simplified model for {\it E. coli} swimming near a solid surface. The mobilities are calculated in Appendix~\ref{bacteria:mob}.\label{summary_approx}}
\end{table}

Now, substituting Eq.~\eqref{Ox_approx} and Eq.~\eqref{Oz_approx} into Eq.~\eqref{Uz_mid} and Eq.~\eqref{Ux_mid} and keeping leading-order terms leads to 
\begin{subeqnarray}
\slabel{Uz_approx}
U_z & \approx & -\frac{\VLU_{xy}\MFO_{zx}\MFO_{yy}}{\NFU_{zz}\MLO_{xx}(\MFU_{yy}+\NFU_{yy})}\,\omega,\\
\label{Ux_approx} U_x & \approx & 
\frac{\WLO_{zy}}
{\displaystyle \left(\frac{\MLO_{zz}(\MFU_{xx}+\NFU_{xx})}{\MFO_{xz}}-\MLU_{zx}\right)}\,\omega,
\end{subeqnarray}
and we get that $U_x>0$ and, more important, that $U_z<0$. This result, together with the result that
$\O_x < 0$, shows that hydrodynamic interactions ``trap"  the cell close to the wall. Note that this trapping does not require  cells to be non-spherical.
Note also that
\begin{equation}
\frac{U_x}{U_y}\approx
\frac{\WLO_{zy}(\MFU_{yy}+\NFU_{yy})}
{\MFO_{yy} \displaystyle \left(\frac{\MLO_{zz}(\MFU_{xx}+\NFU_{xx})}{\MFO_{xz}}-\MLU_{zx}\right)} \approx \frac{3}{\epsilon}\J
\ll 1,
\end{equation}
where $\epsilon={2\pi b }/{\lambda}$  and $\J$ is defined in Appendix~\ref{bacteria:mob}, and
\begin{equation}
\frac{U_z}{U_y}\approx \frac{\VLU_{xy}\MFO_{zx}}{\NFU_{zz}\MLO_{xx}}
\ll \frac{h}{\L} \ll 1,
\end{equation}
so the calculation assumptions  are consistent.

We can finally evaluate the approximate solution for the radius of curvature of the circular trajectory. It is given by
\begin{equation}\label{R_approx}
{\cal R} = \frac{U}{|\O_z|}\approx \frac{U_y}{|\O_z|}\approx \frac{\MLO_{zz}\MFO_{yy}}{\WLO_{zy}(\MFU_{yy}+\NFU_{yy})}
\left(1-\frac{\MFO_{xz}\MLU_{zx}}{\MLO_{zz}(\MFU_{xx}+\NFU_{xx})}\right)
\end{equation}
which is very  similar to that given by the simple physical picture,  Eq.~\eqref{R_model}. 

The results of the analytical model are summarized in Table~\ref{summary_approx}. When we set $\V=\W=0$, and assume that the previous approximations still hold, the results from Appendix~\ref{nowall} (swimming far from surface) are recovered.

\section{Results of the model and comparison with experiments}
\label{bacteria:comparison}

\begin{figure}[t]
\centering
\includegraphics[width=.99\textwidth]{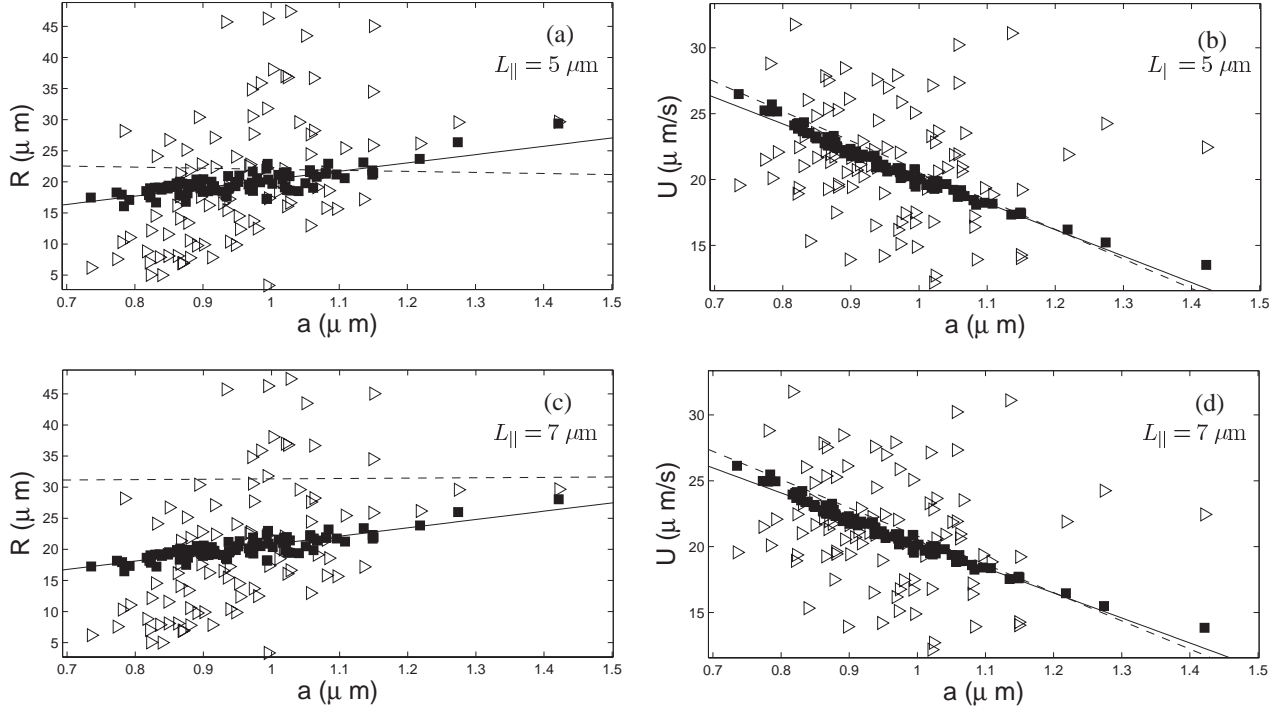}
\caption{Comparison between the results of the experiments ($\triangleright$), the full hydrodynamic model (Eq.~\ref{tosolve}, $\blacksquare$, and best fit, straight line) and the simplified model (Table~\ref{summary_approx}, dashed line). Top:  $\L=5$~$\mu$m, and the values leading to the best least-square fits between the experiments and the full hydrodynamic model, $h=38$~nm and $\omega=211$~Hz:  (a) Radius of curvature, ${\cal R}$,  and (b) swimming velocity, $U$,  as a function of the bacterial radius $a$. Bottom: Same as top  but for $\L=7$~$\mu$m, $h=10$~nm and $\omega=194$ Hz: (c) Radius of curvature,  and (d) swimming velocity  as a function of the bacteria radius $a$.} 
\label{comparisonL}
\end{figure}

We present in this section the comparison between the results of our hydrodynamic model and our experiments. The geometric characteristics of the flagellar bundles that we use are $r= 20$~nm, $\lambda= 2.5$~$\mu$m and $b= 250$~nm \cite{macnab77,magariyama01}. The length of the flagella is usually in the range from $\L= 5$ to 7~$\mu$m, and we will test both values. For the cell radius $a$, we take the equivalent sphere radius $a$ that has the same viscous resistance as the prolate ellipsoid of measured cell dimensions translating along its axis of symmetry \cite{happel}; we use however the measured value of the minor axis of the elliptical head to correctly estimate the distance between the attached helical flagella  and the wall and this results in a small scatter in the theoretical predictions  ($\blacksquare$) of Fig.~\ref{comparisonL}. The only parameter in the model whose value is unknown is the gap thickness $h$. The minimum distance cells can swim from the surface is about 10~nm because of the protrusion of the flagellar hook from the cell body.  Values of $h$ have been measured to be $30-40$~nm \cite{vigeant02} . In order to compare the model with our experimental data, we will assume $h$ to be in the range from 10 to 50~nm.

Despite the large scatter in our experimental data, we find that the results of our hydrodynamic model are consistent with our data, both for the value of the radius of curvature of the trajectory, ${\cal R}$, and the swimming speed, $U=(U_x^2+U_y^2)^{1/2}$. Typical results comparing experiment and theory are illustrated in Fig.~\ref{comparisonL} for $\L=5$~$\mu$m and 7~$\mu$m. The values of the distance to the wall $h$ and the flagella rotation speed $\omega$ were chosen to lead to the best fit of radius of curvature and velocity of the full hydrodynamic model (square symbols and straight line) to the experimental data (triangular symbols); we find $h=38$~nm and $\omega=211$~Hz for $\L=5$~$\mu$m, and $h=10$~nm and $\omega=194$~Hz for $\L=7$~$\mu$m. These values are consistent with the measurements of \cite{vigeant02} and with typical values for the rotation rate of flagella in {\it E. coli} \cite{lowe87}.

\begin{figure}[t]
\centering
\includegraphics[width=.99\textwidth]{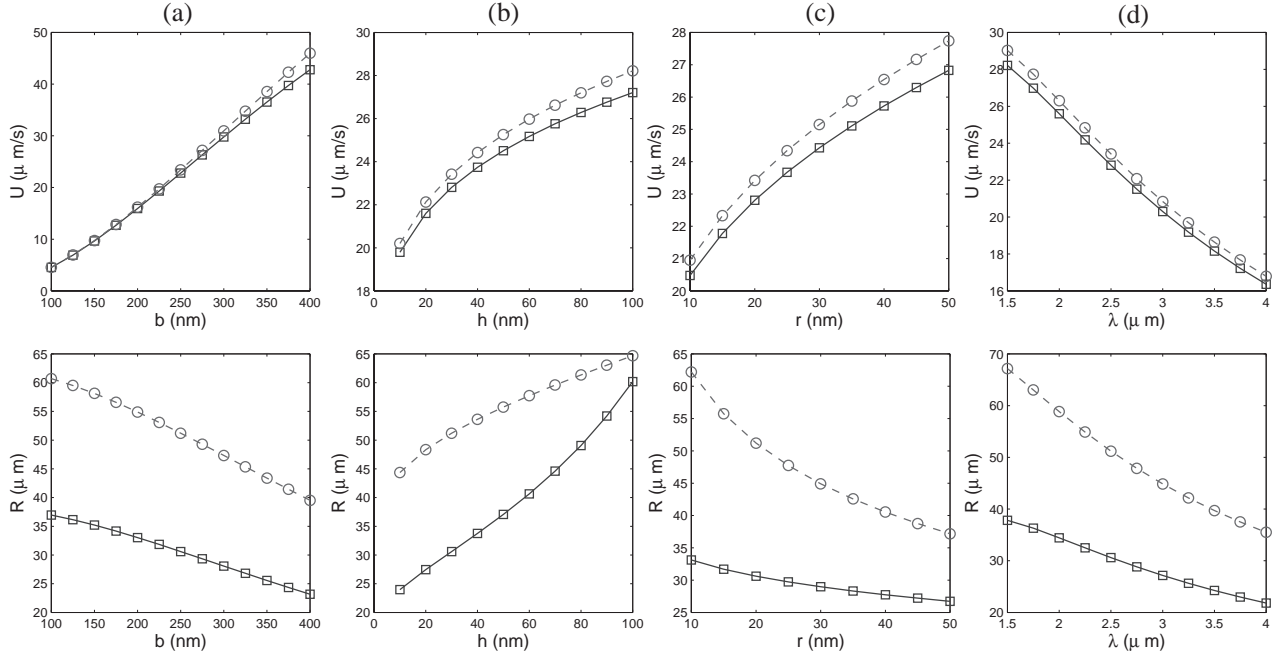}
\caption{Dependence of the results $\{U,{\cal R} \}$   on the geometrical parameters $\{b,h,r,\lambda\}$ for the two models (full model: squares and straight line; approximate analytical model: circles and dashed line), in the case where $r= 20$~nm, $\lambda= 2.5$~$\mu$m, $b= 250$~nm, $\L=7$~$\mu$m, $h=30$~nm, $a=1$~$\mu$m and $\omega=200$~Hz, and one of the parameters is varied at a time. Dependence on (a) the helix radius $b$, (b) the gap thickness $h$, (c) the helix half-thickness $r$, and (d) the helix wavelength, $\lambda$. } 
\label{variations}
\end{figure}

Furthermore, we observe in Fig.~\ref{comparisonL} that the approximate analytical model (Table~\ref{summary_approx}, dashed line) gives a good approximation of the swimming speed for both values of $\L$ but fails to predict  the correct radius of curvature of the trajectory for $\L=7$~$\mu$m (the dashed line appears however to go to through some of the experimental data due to the large scatter in the measured radius of curvature). More importantly, the approximate analytical model fails to capture the increase of the radius of curvature of the trajectory with the cell size, an experimental feature that is predicted by the full hydrodynamic model.

Let us now investigate the dependence of the models, that is the values of $\{{\cal R},U\}$ given by both the full model and the approximate analytical model, with other geometrical parameters describing the bacteria:  $\{b,h,r,\lambda\}$\footnote{The dependence of the model on the cell body, $a$, is illustrated in Fig.~\ref{comparisonL}. Moreover, from Eq.~\eqref{tosolve}, it is straightforward to see that both ${\bf U}$ and $\boldsymbol{\Omega}$ scale with $\omega$, and therefore $\cal R$ is independent of $\omega$.}. In order to display the variations, we will fix the values to be $r= 20$~nm, $\lambda= 2.5$~$\mu$m, $b= 250$~nm, $\L=7$~$\mu$m, $h=30$~nm, $a=1$~$\mu$m and $\omega=200$~Hz, and will then vary one of these parameters at a time. The results are displayed in Fig.~\ref{variations} for the full hydrodynamic model (solid lines, squares) and the approximate analytical model (dashed line, circles).

\begin{figure}[t]
\centering
\includegraphics[width=.6\textwidth]{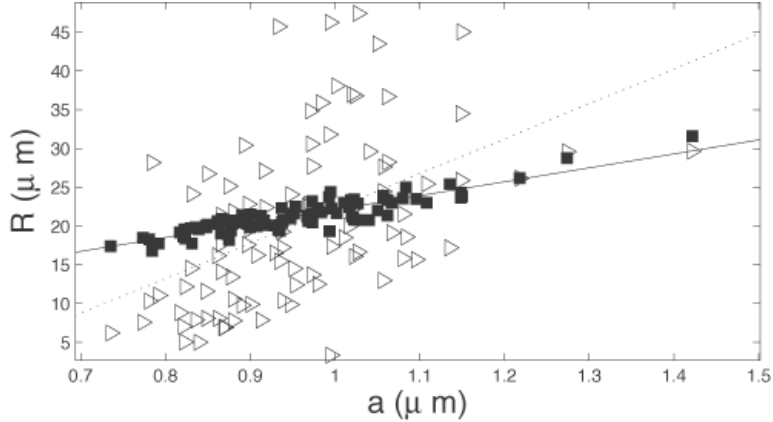}
\caption{Best fit  to the experimental data ($\triangleright$) by a $h(a)$ law for $\L=7$~$\mu$m and $\omega=194$ Hz, as given by Eq.~\eqref{fiteq} ($\blacksquare$ and best fit, straight line). The dotted line represent the best fit of the experimental data by a straight line (in the least-square sense), which we have added as a guide to the eyes.} 
\label{bestfit}
\end{figure}

These results first confirm that both models are in agreement for the trends and values of the swimming velocity, but that the approximate analytical model overestimates the value of the radius of curvature of the trajectory (by up to 50\%). The dependence of the swimming velocity,  $U$, is found to be consistent with the increase of the propulsive viscous force with $b/\lambda$ and $r$, and the decrease of the viscous resistance with $h$ (see the values of the mobilities as calculated in Appendix \ref{bacteria:mob}). The radius of curvature increases with $h$, consistent with a decreases of the induced $z$-torque with the distance to the surface. The radius also decreases with $r$, consistent with an increase in the hydrodynamic interactions with the nearby surface as described by Eq.~\eqref{drag}. Furthermore, $\cal R$ decreases with $b$, confirming the important role of the viscous resistance on parts of the helix that are close to the surface (whose distance to the surface decreases with $b$) in inducing the torque on the cell in the $z$-direction.

Finally, we note that the value of the radius of curvature from the model depends strongly on the unknown gap thickness $h$. Returning to the comparison with the results of our experiments, we see that data for larger cells tend to be more consistent with the model for large values of $h$ (Figs.~\ref{comparisonL} and \ref{variations}). Thus we propose here that, if we suppose that all bacteria have the same geometrical characteristics, $\{b,r,\lambda,\L\}$, 
our hydrodynamic model could be used to estimate the relation between the typical cell size, $a$, and its steady-state distance to the wall, $h$, that is, the $h(a)$ relationship. The results are illustrated in Fig.~\ref{bestfit}  where we have plotted together the results of the experiments with the predictions of the hydrodynamic model with a linear relationship between $a$ and $h$,
\begin{equation}\label{fiteq}
h(a)=h_0 + \left(\frac{a-a_1}{a_2}\right)h_1,
\end{equation}
which leads to the best overall fit to the experimental data;  the parameters for this fit are $h_0=10$~nm, $a_1=0.7$~$\mu$m, $a_2=0.8$~$\mu$m and $h_1=11$~nm. 

\section{Conclusion}

We have presented  a hydrodynamic model for the swimming of {\it E. coli} near solid boundaries and compared it to a new set of measurements of cell velocities and trajectories. We have shown that force-free and torque-free swimming was responsible for the clockwise circular motion of the cells, as well as for their hydrodynamic ``trapping''  close to the surface, that is, $\Omega_x<0$ and $U_z<0$. The fact that cells do not eventually come in contact with the surface is probably due to other short-range surface-cell interactions \cite{vigeant97,vigeant02}.

The main assumptions made in this paper, and which illustrate the differences between real swimming {\it E. coli} cells and our model, are the following: (1) We have replaced the bundle of several flagella by a single  rigid helix; (2) We have assumed that the the cell body was spherical; (3) We have ignored all interactions between the cell body and the flagella. Relaxing these assumptions would improve on the agreement between theory and experiments, but we do not expect it would change the physical picture given in this paper. Including the presence of a second (top) boundary should also modify the cell trajectories. 

Finally, we wish to remark that if the surface was a perfectly-slipping interface (such as the free surface between air and water) instead of a no-slip surface, the change of the direction of the image system for a point force \cite{blake71} should lead to bacteria swimming in circles, but in a counterclockwise direction (X. L. Lu, University of Pittsburgh, private communication).

\section*{Acknowledgements}
We thank H. Berg for his gift of strain HCB437 as well as H. Berg, H. Chen and L. Turner for helpful conversations and feedback.   This work was supported by the NIH (GM065354), DOE (DE-FG02-OOER45852) and the Harvard MRSEC.  WRD acknowledges an NSF-IGERT Biomechanics Training Grant (DGE-0221682). EL acknowledges funding by the Office of Naval Research (Grant \# N00014-03-1-0376).


\appendix

\section{Cell mobilities}
\label{bacteria:mob}

We present in this Appendix the values of the hydrodynamic mobilities of the bacteria. First, since we have $h\ll a$, the lubrication approximation can be made to derive the mobilities for the 
cell body \cite{jeffrey,goldman67,oneill67,cooley68}. We find that they are given by 
\begin{subeqnarray}
\NFU_{xx} & = & \displaystyle \NFU_{yy} = 6\pi\mu a \left[ \frac{8}{15}\ln\left(\frac{a}{h}\right)+0.96\right], \\
\NFU_{zz} & = &\displaystyle 6\pi \mu \frac{a^2}{h},\\
\NLO_{xx} & = &\displaystyle \NLO_{yy} = 8\pi\mu a^3 \left[ \frac{2}{5}\ln\left(\frac{a}{h}\right) +0.38\right],\\
\NLO_{zz} & = &\displaystyle 8\pi\mu a^3\\
\VLU_{xy} & = &\displaystyle \VLU_{yx} = 8\pi\mu a^2\left[\frac{1}{10}\ln\left(\frac{a}{h}\right)-0.19 \right],\\
\VFO_{xy} & = &\displaystyle \VFO_{yx} = 6\pi \mu a^2 \left[\frac{2}{15}\ln\left(\frac{a}{h}\right)-0.25 \right],
\end{subeqnarray}
Note that we  assumed that $\NLO_{zz}$  was equal to its far-field value, as it was shown that the presence of a nearby surface has only a small effect on the value of this mobility \cite{jeffrey}.

Second, the bundle of helical flagella is described by the equation
\begin{equation}
\left\{
\begin{array}{l}
 x  =  b \sin (s-\omega t) \\
y  = \displaystyle -\frac{\lambda}{2\pi}(s+s_0) \\
z  =  b \cos (s-\omega t)
\end{array}
\right.
\end{equation}
where $s$ ranges from $0$ to $2n\pi$, and $\omega$ is the rotation rate of the flagella bundle relative to the cell body. In that case, the mobility calculation was done according to the principle introduced in \S \ref{modeling} and we get
\begin{subeqnarray}
\MFU_{xx} & = &\displaystyle \MFU_{zz} = 2\, \c\, \L\, \frac{1 + 3 \e^2/4}{(1+\e^2)^{1/2}},\\
\MFU_{yy}& = &\displaystyle \c\,\L\, \frac{1 + 2\e^2\,\,\,\,}{(1+\e^2)^{1/2}}, \\
\MLU_{yy} & = &\displaystyle\MFO_{yy}= \c\, b\,\L\, \frac{\e}{(1+\e^2)^{1/2}}, \\
\MLO_{xx}& = &\displaystyle \MLO_{zz} = \frac{2}{3}\,\c\, \L^3\,  \frac{1 + 3 \e^2/4}{(1+\e^2)^{1/2}},\\
\MFO_{zx}& = &\displaystyle \MFO_{xz} = \MLU_{xz}  =  \MLU_{zx} = \c\,\L^2\, \frac{1 + 3 \e^2/4}{(1+\e^2)^{1/2}}, \\
\MLO_{yy}& = &\displaystyle 2\,\c\, b^2\, \L\, \frac{1 + \e^2/2}{(1+\e^2)^{1/2}}, \\
\MLU_{xx} & = &\displaystyle \MLU_{zz} = \MFO_{xx}= \MFO_{zz}=\frac{1}{2}\c\, b\,\L\, \frac{\e}{(1+\e^2)^{1/2}} ,\\ 
\WLU_{yx}& = &\displaystyle  \WFO_{xy} = -2\,\c\, b\,\L\,  \frac{1 + \e^2/2}{(1+\e^2)^{1/2}} \I ,\\
\WFU_{xy}& = &\displaystyle\WFU_{yx}= -\c\,\L\, \frac{\e}{(1+\e^2)^{1/2}} \I ,\\
\WLU_{zy }& = &\displaystyle \WFO_{yz}=-\c\, \L^2\, \frac{\e}{(1+\e^2)^{1/2}} \J, \\
\WLU_{xy}& = &\displaystyle \WFO_{yx}=-\c \,b \,\L\, \frac{1 + 2\e^2\,\,\,\,}{(1+\e^2)^{1/2}}  \I ,\\
\WLO_{zy}& = &\displaystyle \WLO_{yz}=-2 \,\c\, b\, \L^2\, \frac{1 + \e^2/2}{(1+\e^2)^{1/2}}\J, \\
\WLO_{xy}  & = &\displaystyle \WLO_{yx} = -\c\,b^2\,\L\, \frac{\e}{(1+\e^2)^{1/2}} \I, 
\end{subeqnarray}
where $\epsilon = 2 \pi  b / \lambda$, and where we have defined the two integrals
\begin{equation}
\I=\int_0^1 \cos (2\pi u ) f(\cos(2\pi u) ) {\rm d} u,\quad \J=  \int_0^1 (u + u_0)\cos (2\pi n u) f (\cos(2\pi n u )) \d u.
\end{equation}
Note that for the calculation of $\MLO_{yy}$, the contribution due to the local rotation of the 
flagella can be neglected because $r\ll b$ \cite{chwang71}.

\section{Swimming far from a surface}\label{nowall}
When the bacteria swims away from a surface, we have $\W=0$ and $\V=0$, so the mobility matrices become
\begin{equation}
{\cal A}=
\left(
\begin{array}{cccccc}
 -\NFU_{xx}  &  0   &   0   & 0  & 0  & 0 \\
 0   &  -\NFU_{yy}   &   0   & 0   & 0  & 0 \\
 0   &  0   &   -\NFU_{zz}    & 0  & 0  & 0 \\
 0   &  0   &  0   & -\NLO_{xx}  & 0  & 0 \\
0    &  0   &  0   & 0  & -\NLO_{yy}  & 0 \\
 0   &  0   &  0   & 0  & 0  & -\NLO_{zz} \\
\end{array}
\right),
\end{equation}
and
\begin{equation}
{\cal B}= \left(
\begin{array}{cccccc}
 -\MFU_{xx}  &  0  &   0   & \MFO_{xx}  & 0  & -\MFO_{xz} \\
0   &  -\MFU_{yy}  &   0   & 0  & -\MFO_{yy}   & 0 \\
 0   &  0   &  -\MFU_{zz}   & \MFO_{zx}   & 0  & \MFO_{zz} \\
 \MLU_{xx}   &  0   &  \MLU_{xz}  & -\MLO_{xx}  & 0  & 0 \\
0 &  \MLU_{yy}    &  0   & 0 & -\MLO_{yy}  & 0 \\
  -\MLU_{zx}  &  0   &  \MLU_{zz}   & 0  & 0   & -\MLO_{zz} \\
\end{array}
\right)\cdot
\end{equation}
Solving Eq.~\eqref{tosolve} for the velocities and rotation rates in this case leads to 
 $U_x=U_z=\O_x=\O_z=0$ and 
\begin{subeqnarray}
U_y & = & \frac{\MFO_{yy}\NLO_{yy}}{(\MLO_{yy}+\NLO_{yy})(\MFU_{yy}+\NFU_{yy})+\MFO_{yy}\MLU_{yy}} \, \omega\\
\O_y & = & \frac{\MLO_{yy}(\MFU_{yy}+\NFU_{yy})+\MFO_{yy}\MLU_{yy}}{(\MLO_{yy}+\NLO_{yy})(\MFU_{yy}+\NFU_{yy})+\MFO_{yy}\MLU_{yy}} \, \omega. 
\end{subeqnarray}
In the absence of a wall, the bacteria swims therefore in a straight line and rotates its body in the direction opposed to that of the flagella.

\bibliographystyle{unsrt}
\bibliography{bib_thesis_elauga}

\end{document}